# Damping of micromechanical structures by paramagnetic relaxation

J.G.E. Harris [a], R. Knobel [a], K.D. Maranowski [b], A.C. Gossard [b], N. Samarth [c], and D.D. Awschalom [a),b),d)]

*Center for Spintronics and Quantum Computation, University of California, Santa Barbara, California, 93106*

Abstract

We find that the damping of micromechanical cantilevers is sensitive to the relaxation dynamics of paramagnetic ions contained within the levers. We measure cantilevers containing paramagnetic Mn ions as a function of temperature, magnetic field, and the vibrational mode of the lever and find that the levers' damping is strongly enhanced by the interplay between the motion of the lever, the ions' magnetic anisotropy, and the ratio of the ions' longitudinal relaxation rate to the resonance frequency of the cantilever. This enhancement can improve the levers' ability to probe the relaxation behavior of paramagnetic or superparamagetic systems; it may also represent a previously unrecognized source of "intrinsic" dissipation in micromechanical structures.

a) Department of Physics, University of California, Santa Barbara, CA 93106
b) Department of Electrical and Computer Engineering, University of California, Santa Barbara, CA  93106
c) Department of Physics, Pennsylvania State University, University Park, PA 16802
d) Contact email: awsch@physics.ucsb.edu

The damping of a micromechanical structure reflects the flow of energy from the structure's macroscopic motion into the microscopic degrees of freedom of its environment. This dissipation can provide quantitative information about the response of the environment to the motion of the structure. Electrostatic coupling,[1,2] magnetic domain wall motion,[3,4] thermoelastic effects,[5] and a number of other phenomena have been probed with exceptional sensitivity and/or spatial resolution by measuring their effect upon the dissipation of micromechanical structures. However, this dissipation also limits the ultimate sensitivity of measurements which use the narrow resonance of micromechanical structures to probe conservative forces (e.g., magnetic resonance force microscopy[6], torsional magnetometry[4] and some types of atomic force microscopy[7]). At present the dissipation mechanisms which limit such measurements are not well understood.

In this paper we demonstrate that a micromechanical cantilever's damping can provide a sensitive probe of the relaxation dynamics of paramagnetic ions contained within the lever. We measure the damping of cantilevers containing paramagnetic ions as a function of temperature, applied magnetic field, and the vibrational mode of the lever. We find that our results can be explained by a simple model in which energy from the macroscopic motion of the lever is coupled to the ions' magnetic anisotropy and then dissipated by their longitudinal relaxation. The strength of this effect depends primarily upon the ions' anisotropy and the ratio of their relaxation rate to the lever's resonance frequency.

To the best of our knowledge the connection between paramagnetic relaxation and the damping of micromechanical structures has not previously been studied. In addition to providing insight into the relaxation dynamics of small magnetic systems (of importance to their use in data storage or spintronic applications), the damping resulting from this effect may play an important role in limiting the ultimate sensitivity of micromechanical oscillator measurements.

The paramagnetic ions studied here are $Mn^{2+}$ in a $Zn_{0.87}Cd_{0.13}Se$ host. The spin-5/2 Mn ions are isoelectronic with their host material, and at the concentration of these samples (~ 1 %) are known to be paramagnetic over the relevant temperature range.[8] The epilayers which form both the magnetic samples and the cantilevers are grown by molecular beam epitaxy (MBE) as a single hybrid III-V/II-VI heterostructure (Fig. 1(a)).

The heterostructure is patterned into free-standing 100 nm thick GaAs cantilevers, each of which supports a mesa of the magnetic material. The fabrication of these integrated cantilever/sample structures has been described in detail previously.[9]

Figure 1(b) is an SEM photo showing a cantilever of length $L$ = 250 μm supporting a 40 μm x 100 μm rectangular mesa of the paramagnetic II-VI material. The levers used in the measurements described here are nominally identical except that the sample mesas are 15 μm-radius discs (containing ~ 2 x $10^9$ $Mn^{2+}$ ions). Measurements were performed with the cantilevers mounted in a sorption-pumped $^3$He cryostat[10] over a range of temperature 0.32 K < $T$ < 10 K. A magnetic field $H$ was applied normal (+/- ~ 3°) to the plane of the sample. A piezoelectric crystal was used to drive the cantilever with an amplitude $x$, which was measured by an optical fiber interferometer. The interferometer used a 1310 nm laser with < 30 nW of power to minimize unintentional heating of the sample. For $T$ < 10 K, the lowest flexural resonance of the cantilever was $\omega_1/2\pi$ = 835 Hz. Similar structures were used in earlier work to measure the equilibrium magnetization $M$ of integrated samples.[9,11] Here we focus exclusively on the damping of the cantilever.

The damping of the cantilever is characterized by the energy it dissipates in each period of its motion, $\Delta E_1 = \pi k_1 x^2/Q_1$. Here $k_1$ is the spring constant of the lever's lowest flexural mode (calculated from the dimensions of the lever to be 7 x $10^{-5}$ N/m) and $Q_1$ is the quality factor of the lowest flexural mode. We use two methods to determine $\Delta E_1$: in the first we drive the cantilever at $\omega_1$ using a phase-locked loop and vary the strength of the drive to maintain constant $x$ (~ 100 nm). This drive is proportional to $\Delta E_1$, but does not provide an absolute value. In the second method we measure the resonance curve of the cantilever and fit the data to extract $Q_1$. This approach is more time consuming, but provides an absolute value of $\Delta E_1$.

Fig. 2(a) shows $\Delta E_1$ as a function of $T$ for different values of $H$. The data shown as lines were taken using the first of these methods as the temperature of the cryostat was gradually changed. The absolute scale of each curve in Fig. 2(a) was then determined by measuring the resonance curve of the lever at several different values of $H$ and $T$, shown as solid points in Fig. 2(a). One such resonance curve and its fit are shown in the inset of Fig. 2(a).

The data show a strong temperature dependence down to 0.32 K, indicating that thermal contact is maintained between the cantilever and cryostat. At temperatures below 2 K, the presence of a magnetic field produces a dramatic increase in $\Delta E_1$. For $H = 2$ T, $\Delta E_1$ is peaked near $T = 0.6$ K. This is in marked contrast to the data for $H = 0$ T, in which $\Delta E_1$ varies by only ~ 2 % over the same temperature range.

The data for $H = 0$ T are similar to those from cantilevers which do not contain paramagnetic ions, and correspond to the lever's intrinsic (non-magnetic) mechanical dissipation. The increase of $\Delta E_1$ for $H > 0$ T is only observed in levers containing paramagnetic ions. In order to understand the origin of this additional damping, we note that paramagnets in an oscillating magnetic field produce dissipation when their longitudinal relaxation rate $1/T_1$ and the frequency of the oscillating field are comparable. For $Mn^{2+}$ in II-VI host materials, $T_1$ increases dramatically with decreasing $T$, and data[12] for $T > 2$ K suggest that in the temperature range shown in Fig. 2(a), $1/T_1$ should approach $\omega_1$, thereby making it possible for the motion of the lever to produce dissipation via the paramagnetic ions.

In our setup the oscillation of the lever in the static, uniform magnetic field $H$ is equivalent to an oscillating magnetic field because of the non-zero (but weak)[13] magnetic anisotropy of the $Mn^{2+}$ ions. This can be understood as follows: in a static applied field $H$ the six Zeeman levels of the $Mn^{2+}$ ion are each split by $E_z \sim g\mu_B H$, where $g$ is the Landé $g$-factor for $Mn^{2+}$ and $\mu_B$ is the Bohr magneton. However, the presence of any magnetic anisotropy in the sample will cause $E_z$ to depend also upon the angle $\theta$ between $H$ and the sample.[14] Thus as the sample rotates with each oscillation of the lever,[15] the splitting of the Zeeman levels will change by an amount $\Delta E_z \approx \theta_1 \partial E_z/\partial \theta$, where $\theta_1 = 1.377\, x/L \sim 0.5$ mrad is the amplitude of the rotation of the sample. This is formally equivalent to an oscillating magnetic field parallel to $H$ of magnitude $H_{ac}^{eff} = \Delta E_z/g\mu_B$.

In order to compare this model with the data in Fig. 2(a) we note that for weakly anisotropic paramagnetic ions with a single relaxation time $T_1$ in an oscillating magnetic field $H_{ac}$ of frequency $\omega$ superposed on a parallel static field $H$, the energy dissipated per period of the oscillating field is given by[16]

$$\Delta E = \pi H_{ac}^2 \chi_{dc} \frac{\omega T_1}{1+\omega^2 T_1^2}. \tag{1}$$

Previous measurements have shown that in these materials the dc magnetic susceptibility is given by $\chi_{dc} = \partial M/\partial H \propto \partial B_{5/2}(y)/\partial H$ where $B_{5/2}$ is the spin-5/2 Brillouin function. The argument of the Brillouin function is $y = 5g\mu_B H/2k_B T_{eff}$, and the effective temperature $T_{eff}(T)$ is somewhat higher than the physical temperature and reflects the weak interactions between Mn ions.[17] If as a first approximation we make the assumption that $\chi_{dc}$ is the only quantity in Eqn. (1) which varies appreciably with $T$ or $H$, then the temperature dependence of each curve in Fig. 2(a) is simply that of $\chi_{dc}$. In this case the $H$ = 0.4 T and $H$ = 1.0 T data can be understood as reflecting the condition $y \delta 1$. When this condition holds $\chi_{dc}(T)$ and hence $\Delta E_1(T)$ exhibit Curie-like behavior (i.e., both increase roughly as $1/T$), as seen in the data of Fig. 2(a). For $H = 2$ T, this condition is violated and as $T$ decreases $\chi_{dc}(T)$ and hence $\Delta E_1(T)$ reach a maximum and then fall off, reflecting the saturation of the paramagnetic spins as $y >> 1$.

This dependence of $\Delta E(T)$ upon $\omega$ results from the Drude-like term in Eq. (1). This term depends only upon the quantity $\omega T_1(T)$; because $T_1(T)$ decreases with increasing $T$, an increase in $\omega$ means that a given value of $\omega T_1(T)$ occurs at higher $T$. This shifts the overall curve of $\Delta E(T)$ to higher $T$ when $\omega$ is increased from $\omega_1$ to $\omega_2$, as is observed in Figs. 2(a) & 2(b). Figure 3 illustrates this effect over a broader range of temperature for $H = 2$ Tesla.[18]

From this analysis we can conclude that in general the dissipation effects studied here are strongest when the following conditions hold: $\omega T_1$ is of order unity, $\chi_{dc}$ is large, and the anisotropy (and hence $H_{ac}^{eff}$) is also large. In our measurements, the long relaxation time of the $Mn^{2+}$ ions ensures that the first condition is approached at low temperature. However $\chi_{dc}$ in these ions is merely typical for paramagnets, and their anisotropy is notably weak[13]. Given the dramatic impact upon the lever's dissipation observed here, the presence of even a small number of paramagnetic moments in a micromechanical structure (arising from unintentional impurities, broken bonds, or nuclear spins) may be a dominant source of dissipation. We note that the strong magnetic field- and temperature-dependence observed[4,19] in the dissipation of nominally non-magnetic cantilevers may arise from these effects.

The ultimate sensitivity of a cantilever as a detector of paramagnetic relaxation is set by the minimum resolvable dissipation. In our current setup, the intrinsic (nonmagnetic) dissipation of the levers is typically $\Delta E_1(H = 0) \sim 2 \times 10^{-22}$ J (Figs. 2(a) and 3) with changes of 1% resolvable in a 1-second measurement, corresponding to a sensitivity of $2 \times 10^{-24}$ J. The present sample produces a maximum signal roughly $10^3$ times larger than this (Figs. 2(a) and 3), implying a sensitivity to $\sim 2 \times 10^6$ $Mn^{2+}$ ions. Because $\Delta E$ scales as the square of the magnetic anisotropy, a suitable choice of ion and host material may improve this sensitivity substantially. This enhancement may also allow this technique to be applied to the study of the relaxation of single magnetic nanoparticles or magnetic molecules, many of which satisfy the requirement of strong anisotropy, and whose relaxation rate is directly related to their crossover to superparamagnetic and quantum tunneling regimes.[20]

This work was supported by grants NSF DMR-0071888 and –0071977, ONR N00014-03-1-0169 and AFOSR F49620-02-1-0038.

.

**Figure Captions**

Fig.1: (a) Schematic of the MBE-grown heterostructure. The 10.5 nm thick magnetic layer is grown as a digital alloy (seven iterations of: 1/16$^{th}$ monolayer of MnSe followed by 4 15/16$^{th}$ monolayer of Zn$_{0.87}$Cd$_{0.13}$Se). (b) SEM photo of a 100 nm thick, 250 μm long cantilever supporting a rectangular mesa of paramagnetic (Zn,Cd,Mn)Se.

Fig.2: (a) $\Delta E_1$, the energy dissipated by the cantilever in each period of its lowest mode, as a function of temperature for various values of the applied magnetic field. Inset: typical resonance curve (open points) and fit to a damped simple harmonic oscillator (solid line). (b) $\Delta E_2$, the energy dissipated by the cantilever in each period of its second mode, as a function of temperature for various values of the applied magnetic field. In both plots the solid lines are taken as the temperature of the cryostat is slowly changed. The solid points (● $H$ = 0 T; ■ $H$ = 0.4 T; ♦ $H$ = 1 T; ▲ $H$ = 2 T) are from fits of the cantilever resonance taken at fixed temperature. The dashed line in (b) is a guide to the eye.

Fig.3: $\Delta E_1$ and $\Delta E_2$ as a function of temperature at $H$ = 2 T. Each data point is acquired with the temperature of the cryostat held constant. Solid symbols are measurements of the drive amplitude required to maintain constant cantilever amplitude; hollow symbols are from fits of the cantilever resonance curve. The lines are guides to the eye.

(a)

| |
|---|
| ZnSe 100 nm capping layer |
| (Zn,Cd,Mn)Se 10.5 nm sample |
| ZnSe 1100 nm buffer layer |
| GaAs 10 nm buffer layer |
| $Al_{0.7}Ga_{0.3}As$ 100 nm etch stop |
| GaAs 100 nm cantilever layer |
| $Al_{0.7}Ga_{0.3}As$ 100 nm etch stop |
| GaAs substrate |

(b)
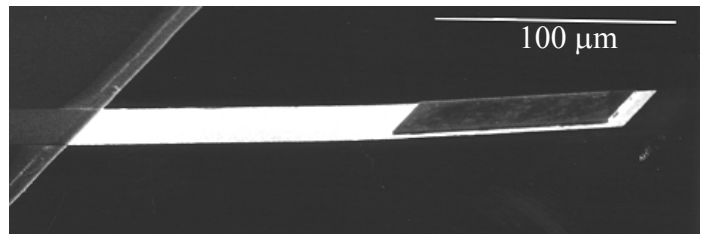

Fig. 1
J.G.E. Harris, *et al*.

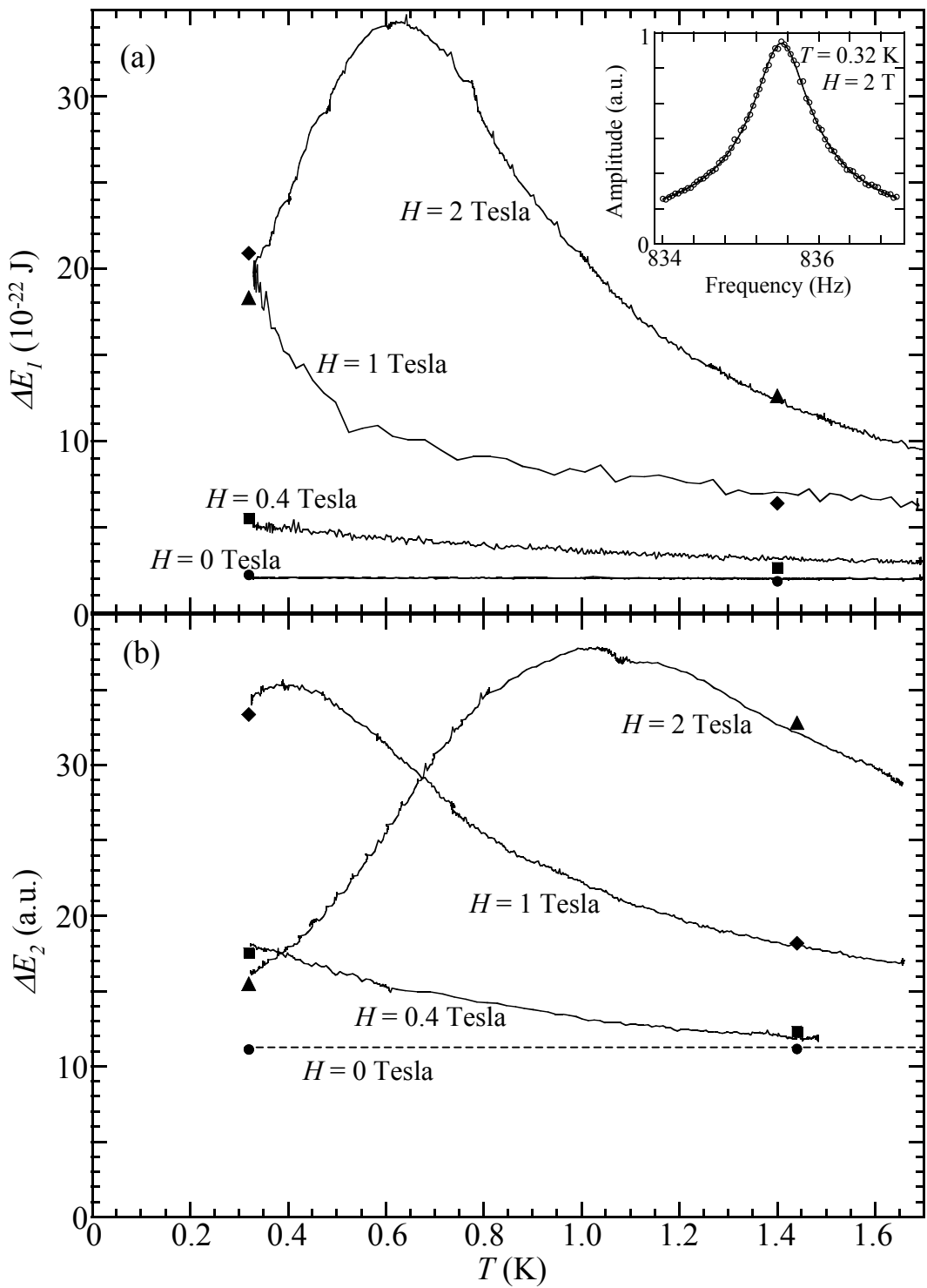

Fig. 2
J.G.E. Harris, *et al*.

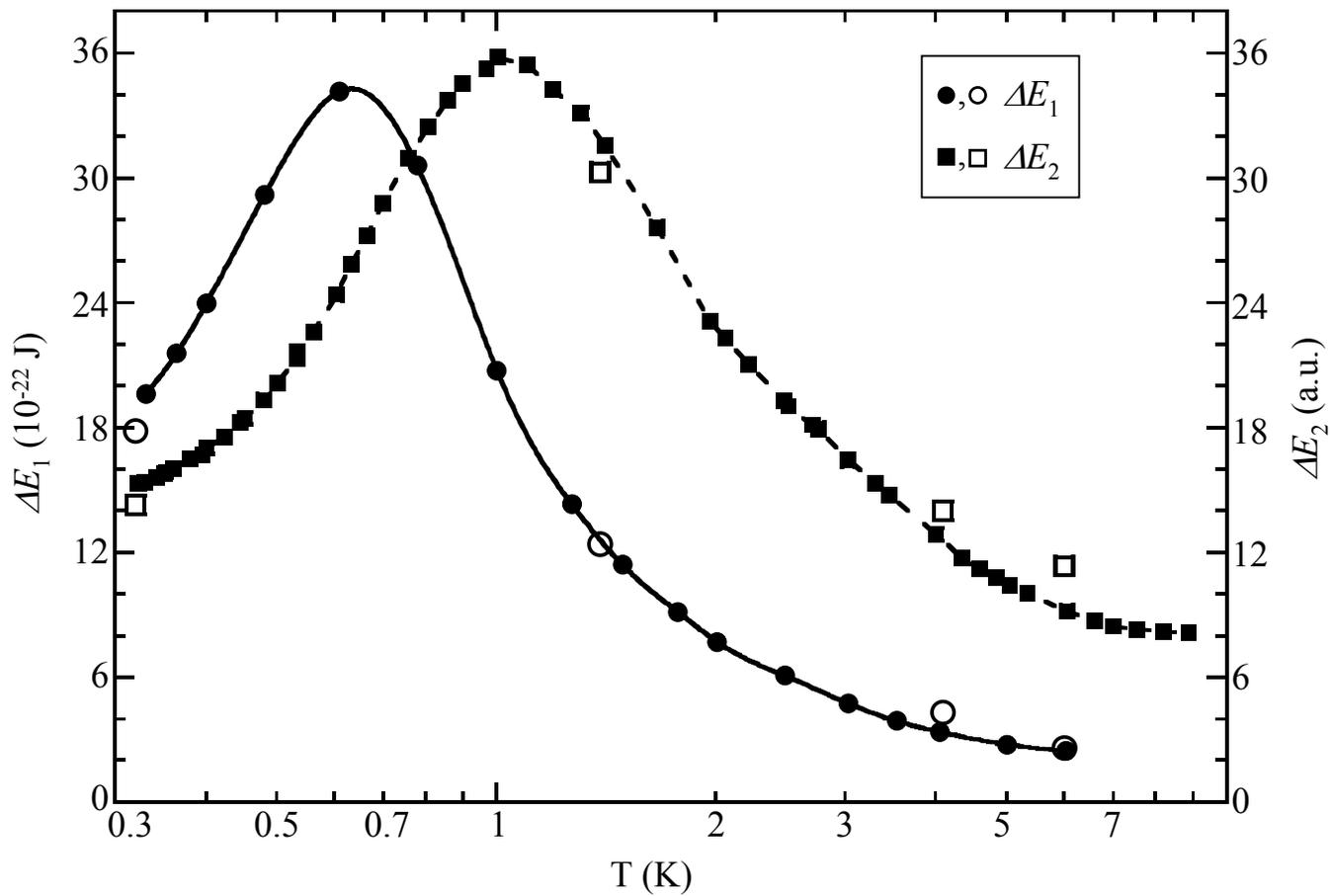

Fig. 3
J.G.E. Harris, *et al*.